\documentclass[12pt,preprint]{aastex}

\shortauthors{Liu et al.}

\begin{document}

\title{Direct In Situ Measurements of a Fast Coronal Mass Ejection and Associated Structures in the Corona} 

\author{Ying D. Liu\altaffilmark{1,2}, Bei Zhu\altaffilmark{3,1}, Hao Ran\altaffilmark{1,2}, Huidong Hu\altaffilmark{1}, Mingzhe Liu\altaffilmark{4,5}, Xiaowei Zhao\altaffilmark{6,7}, Rui Wang\altaffilmark{1}, Michael L. Stevens\altaffilmark{8}, and Stuart D. Bale\altaffilmark{4}}
  
\altaffiltext{1}{State Key Laboratory of Space Weather, National Space Science Center, Chinese Academy of Sciences, Beijing, China; liuxying@swl.ac.cn}

\altaffiltext{2}{University of Chinese Academy of Sciences, Beijing, China}

\altaffiltext{3}{Space Engineering University, Beijing, China}

\altaffiltext{4}{Space Sciences Laboratory, University of California, Berkeley, CA 94720, USA}

\altaffiltext{5}{LESIA, Paris Observatory, PSL University, CNRS, Sorbonne University, University of Paris, France}

\altaffiltext{6}{National Satellite Meteorological Center, China Meteorological Administration, Beijing, China}

\altaffiltext{7}{School of Earth and Space Sciences, Peking University, Beijing, China}

\altaffiltext{8}{Smithsonian Astrophysical Observatory, Cambridge, MA 02138, USA}

\begin{abstract}

We report on the first direct in situ measurements of a fast coronal mass ejection (CME) and shock in the corona, which occurred on 2022 September 5. In situ measurements from the Parker Solar Probe (PSP) spacecraft near perihelion suggest two shocks with the second one decayed, which is consistent with more than one eruptions in coronagraph images. Despite a flank crossing, the measurements indicate unique features of the young ejecta: a plasma much hotter than the ambient medium suggestive of a hot solar source, and a large plasma $\beta$ implying a highly non-force-free state and the importance of thermal pressure gradient for CME acceleration and expansion. Reconstruction of the global coronal magnetic fields shows a long-duration change in the heliospheric current sheet (HCS), and the observed field polarity reversals agree with a more warped HCS configuration. Reconnection signatures are observed inside an HCS crossing as deep as the sonic critical point. As the reconnection occurs in the sub-Alfv\'enic wind, the reconnected flux sunward of the reconnection site can close back to the Sun, which helps balance magnetic flux in the heliosphere. The nature of the sub-Alfv\'enic wind after the HCS crossing as a low Mach-number boundary layer (LMBL) leads to in situ measurements of the near subsonic plasma at a surprisingly large distance. Specifically, an LMBL may provide favorable conditions for the crossings of the sonic critical point in addition to the Alfv\'en surface. 

\end{abstract}


\section{Introduction}

The outer edge of the corona is defined by the Alfv\'en surface, at which the solar wind changes from sub-Alfv\'enic to super-Alfv\'enic \citep{weber1967}. In addition to being sub-Alfv\'enic, the coronal plasma is also characterized by a plasma $\beta$ (the ratio of thermal pressure to magnetic pressure) below 1. The Parker Solar Probe (PSP) mission, launched in 2018 August, was intended to dive below the Alfv\'en surface and make direct in situ measurements of the corona \citep{fox2016, raouafi2023}. PSP measurements indicate that the average value of the Alfv\'en radius is 10 - 12 solar radii from the center of the Sun, but can extend to 20 solar radii or even further for a certain portion of the corona, termed a low Mach-number boundary layer \citep[LMBL;][]{liu2021, liu2023}. An LMBL is a special type of wind emanating from a coronal hole near its boundary along rapidly diverging open magnetic fields. Because of the enhanced Alfv\'en radius of an LMBL, PSP made the first glimpse of the sub-Alfv\'enic plasma for about 5 hr around a heliocentric distance of about 19 solar radii at the eighth encounter on 2021 April 28 \citep{kasper2021}. Now more than 10 steady sub-Alfv\'enic intervals can be identified from PSP measurements since encounter 8, and a statistical analysis of these intervals confirms their nature as LMBLs \citep{jiao2024}. Inside the corona the sonic critical point, at which the solar wind changes from subsonic to supersonic, is usually thought to be well below the Alfv\'en critical point, i.e., a few solar radii \citep[][and references therein]{matthaeus2021}. In situ measurements of the coronal plasma near the sonic critical point have not been reported in the literature so far. 

Loss of stability of the coronal magnetic field can lead to large-scale expulsions of plasma and magnetic flux from the corona \citep[e.g.,][and references therein]{forbes2000}, known as coronal mass ejections (CMEs). CMEs may play a significant role in the evolution of the global coronal magnetic field configuration. \citet{liu2009} provide evidence that the eruption of a CME contributed to long-duration changes in the global coronal field configuration, including a reformed heliospheric current sheet (HCS) that was more warped than before the CME. Their results are consistent with the suggestion of \citet{low1996, low2001} that CMEs are a basic mechanism of coronal magnetic field reconfiguration by removing magnetic flux and helicity from the corona. Magnetic flux carried by CMEs into the heliosphere, on the other hand, would grow without bound in interplanetary space, leading to a problem called ``magnetic field magnitude catastrophe" \citep{gosling1975}. Magnetic reconnection has been invoked to solve the problem \citep[e.g.,][]{mccomas1995, crooker2002}. However, in order to maintain a long-term balance of magnetic flux in the heliosphere reconnection must occur in the sub-Alfv\'enic regime, so reconnected flux can close back to the Sun \citep[e.g.,][]{mccomas1995, crooker2002, deforest2014}. 

When CMEs move into interplanetary space, they are called interplanetary CMEs (ICMEs). ICMEs have been investigated using in situ measurements for more than four decades. In situ signatures of ICMEs include depressed proton temperatures, enhanced helium abundance, bidirectional streaming of electron strahls (BDEs), declining velocity profiles, charge state enhancement, and smooth magnetic fields \citep[e.g.,][]{neugebauer1997, zurbuchen2006}. A subset of ICMEs, termed magnetic clouds (MCs), are characterized by a strong magnetic field, a smooth and coherent rotation of the field, and a depressed proton temperature compared to the ambient solar wind \citep{burlaga1981}. These signatures represent a well-evolved state of CME plasma in interplanetary space. For instance, the depressed temperature results from the expansion of ejecta in the heliosphere \citep{liu2006}, but the signature of charge state enhancement suggests a hot solar source in general \citep{lepri2004}. Also, the low plasma $\beta$ inferred from the depressed temperature and strong magnetic field implies a nearly force-free configuration of the magnetic field \citep[e.g.,][]{goldstein1983, burlaga1988, lepping1990}. However, CMEs cannot be force free at their nascent stage, as forces are needed to launch them from the corona.

Clearly, direct in situ measurements of CMEs and associated structures in the corona are of crucial importance for understanding their nascent state, how they are launched, and their consequences to the corona and heliosphere. A large CME with its driven shock occurred on 2022 September 5 when PSP was near the perihelion of encounter 13. This leads to unprecedented simultaneous imaging and in situ observations of a large CME and shock in the corona. The event has attracted significant attention because of its impressive energetics and in situ measurements in the vicinity of the Sun, for example: \citet{romeo2023} examine the PSP in situ measurements; \citet{patel2023} report white-light observations from PSP/WISPR; \citet{paouris2023} discuss the space weather context of the event; \citet{long2023} look at the solar source region of the eruption. In this work, we provide a different interpretation of the in situ measurements, and identify important features in the measurements that have not been revealed. We show how the young ejecta is compared to well-evolved ICMEs, and illustrate a reforming HCS with clear reconnection signatures as deep as the sonic critical point. This is also a report of in situ measurements of the coronal plasma at the sonic critical point. It suggests a sonic critical radius of about 15 solar radii from the center of the Sun in the present case, which is surprising. We describe coronagraph imaging observations in Section 2 and PSP in situ measurements in Section 3. The conclusions are summarized in Section 4. 

\section{Coronagraph Imaging Observations}

The 2022 September 5 CME is an impressive eruption, which occurred from NOAA AR 13088 (W170$^{\circ}$S28$^{\circ}$). Figure~1 shows coronagraph images of the CME and shock from SOHO and STEREO A, which observed the event from behind (see Figure~2). The CME seems distorted, as indicated by a lobe on the east and another lobe on the southwest (Figure~1a). An alternative interpretation is that there are two CMEs. \citet{long2023} suggest a smaller eruption (corresponding to the lobe on the east) from a source region west of AR 13088 using EUV observations from Solar Orbiter (SolO). The X-ray fluxes from the low-energy channels of STIX aboard SolO show multiple peaks \citep{long2023, patel2023}, which may also imply more than one eruptions.

Given the complexity in the images, modeling of the complex CME (or the components separately) would be difficult. However, the shock, which appears as a faint edge around the complex CME, can be modeled well by a simple spherical structure in both views. The situation is similar to the complex events of 2012 July 23 \citep{liu2017} and of 2017 July 23 \citep{liu2019b}. There are four free parameters in the spherical shock model, the longitude and latitude of the shock propagation direction, the distance of the shock center from the Sun, and the radius of the shock sphere \citep[e.g.,][]{hess2014, kwon2014, kwon2015, liu2017, liu2019a, liu2019b}. The two views from SOHO and STEREO A are used simultaneously to fit the shock (see an example at 16:46 UT in Figure~1). Since only a few images are available from SOHO, we determine the longitude and latitude based on the simultaneous observations from the two spacecraft, and then follow their values and only adjust other parameters for observations of single spacecraft (i.e., STEREO A). The shock modeling yields a propagation direction, which is about $170^{\circ}\pm5^{\circ}$ west and $50^{\circ}\pm10^{\circ}$ south of the Earth. The propagation longitude is consistent with the active region longitude, but the propagation latitude is considerably larger than the source region latitude. This may suggest a deflection of the complex CME in latitude. Indeed, EUV observations from SolO indicate southward propagating coronal waves from the source region \citep{long2023}. The peak speed of the shock nose is $2900\pm200$ km s$^{-1}$ from the modeling. For comparison, \citet{patel2023} obtain an average speed of about 2500 km s$^{-1}$ for the CME leading edge using PSP/WISPR observations. PSP appeared in the field of view of STEREO A during the encounter, which enabled the simultaneous imaging and in situ observations of the event.    

Figure~2 displays the cross section of the modeled shock in the ecliptic plane. The shock expanded to enclose the whole Sun. This must be driven by the vast expansion of the ejecta. Previous cases have shown similar shock geometries with 360$^{\circ}$ envelope around the Sun \citep[e.g.,][]{kwon2015, liu2017, liu2019a, zhu2018, hu2019}. Note that at some point the structure near the wake could be just a wave without a non-linear steepening character, and the shock may quickly decay in the backward direction \citep[see more discussions in][]{liu2017, liu2019a}. Given the propagation direction and PSP location, PSP would encounter the flank of the ejecta and shock. The modeling suggests that at 17:27 UT, when in situ measurements at PSP show a shock passage, the modeled shock does not hit PSP yet. The model predicts shock arrival at PSP at 17:39 UT with a shock normal velocity of $1300\pm 100$ km s$^{-1}$. \citet{paouris2023} obtain a similar shock speed at 17:38 UT, using projected speed measurements from STEREO A along the position angle of PSP. When calculating the shock arrival, we have taken into account the propagation time of photons from the shock to the imaging spacecraft near 1 au (about 8 minutes). The time series of coronagraph images that can be modeled ends at 17:51 UT (from STEREO A), which translates to about 17:43 UT at PSP. This is beyond the shock arrival time at PSP, and we have used interpolation to determine the shock parameters at the two times (17:27 and 17:39 UT).  

\section{PSP In Situ Measurements} 

\subsection{Overview} 

The in situ measurements are made by the FIELDS instrument suite \citep{bale2016} and the SWEAP package \citep{kasper2016} aboard PSP. Ion data (protons and alphas) are from the ion electrostatic analyzer \citep[SPAN-I;][]{livi2022}, and electron data are from the two combined electron electrostatic analyzers \citep[SPAN-E;][]{whittlesey2020, halekas2020}. Electron parameters (including density and core temperature) can also be obtained from quasi-thermal noise (QTN) spectroscopy \citep{moncuquet2020}. The QTN density is considered to be the most reliable, as it is derived from measurements of the local plasma frequency.

Figure~3 shows an overview of the in situ measurements near the perihelion of encounter 13. We first see switchbacks in the data (Figure~3d and 3f), i.e., Alfv\'enic flows with reversed magnetic field and enhanced radial velocity that are prevalent in PSP measurements since the first encounter \citep{bale2019, kasper2019}. The magnetic field deflection angle (Figure~3h) is derived using the method of \citet{liu2023}, with its sign indicating the deflection direction. The radial velocity variation (Figure~3i) is obtained by subtracting a low-pass filtered ``baseline" value from the observed radial velocity \citep{liu2023}. Since the majority of deflection angles are below $90^{\circ}$, \citet{liu2023} suggest that the term ``switchbacks" is better changed to ``Alfv\'enic flows with a deflected magnetic field and enhanced radial velocity," or ``Alfv\'enic deflections" (ADs) for short. We will follow and use the term of ADs in this work. Then we see a shock passage at 17:27 UT on September 5 followed by an ejecta, during which ADs are replaced by other types of fluctuations. They come back after the ejecta but with decreased amplitudes (Figure~3f, 3h and 3i). Their amplitudes are further reduced after the HCS crossing on September 6. Finally, ADs recover to their pre-shock level at the end of September 7. Note that the radial velocity variation is largely one-sided (i.e., enhancement) outside the transient plasma (Figure~3i). This is a typical signature of ADs in PSP encounter measurements. Inside the ejecta the radial velocity variation is two-sided. The few negative values just upstream of the shock are caused by the low-pass filtering that fails to capture the abrupt change at the shock. Also, the field deflection angle is different inside the ejecta, which is more or less coherent. These two features are useful in distinguishing the transient plasma from the ambient wind. 

\citet{liu2023} find a dependence of the amplitudes of ADs on the radial Alfv\'en Mach number, and a low Alfv\'en Mach number will suppress the amplitudes of ADs because of their nature as Alfv\'enic fluctuations. The way that ADs behave in Figure~3 is consistent with their theory. After the ejecta the Alfv\'en Mach number decreases to below 1, and further drops to about 0.1 after the HCS crossing (Figure~3g). We see corresponding changes in the amplitudes of ADs with the Alfv\'en Mach number. The radial component of the magnetic field is very smooth after the HCS crossing, and ADs almost completely disappear. Note that the radial sonic Mach number descends to about 1, which is accompanied by a plasma $\beta$ of 0.01 or lower. The calculation of the plasma $\beta$ includes contributions from protons, electrons and alphas. In calculating the radial sonic Mach number, we use the sound speed defined as $c_{\rm s} = \sqrt{\gamma k_{\rm B}(T_p + T_e)/m_p}$, where $\gamma=5/3$, $k_{\rm B}$ is the Boltzmann constant, $T_p$ the proton temperature, $T_e$ the electron temperature, and $m_p$ the proton mass. The in situ measurements of the coronal plasma deep to the sonic critical point imply a surprisingly large sonic critical radius, which is about 15 solar radii from the center of the Sun. Our magnetic mapping (see below) indicates that the sub-Alfv\'enic intervals around the HCS crossing are LMBLs defined by \citet{liu2023}. Perhaps an LMBL may also enable an easier crossing of the sonic critical point. Based on the measurements near the sonic critical point, we suggest that ADs may start from the sonic critical point, as below it the magnetic field is too strong to be deflected.

When writing this paper, we notice the work by \citet{romeo2023} looking at the same in situ measurements in detail. They put forward pictures of how the in situ measurements may be connected with white-light imaging observations. As can be seen, our interpretation is different from theirs, and we identify important features in the measurements that have not been revealed. For example, \citet{romeo2023} argue that the sub-Alfv\'enic intervals are also transient ejecta plasma (see their Figure~12 and corresponding discussions). Here we suggest that they are the ambient wind using the AD properties as a support. In addition, the velocity is as low as 100 - 200 km s$^{-1}$ around the HCS crossing (Figure~3d). If they were part of the ejecta plasma, the velocity would be much higher considering the impressive energetics of the ejecta in coronagraph images. We do not see other ejecta signatures either for those time periods. The low plasma $\beta$ around the HCS crossing is an indication of the deeper corona, not an ejecta signature. Also note the solar source latitude ($\sim$S28$^{\circ}$) and the propagation latitude of the complex shock ($\sim$S50$^{\circ}$). Both are south of the ecliptic plane. The source region, which may correspond to the smaller eruption as suggested by \citet{long2023}, is also south of the ecliptic plane. It would be difficult for PSP in the ecliptic plane to cross a CME leg when the ejecta is gone. This again supports the interpretation of the data behind the ejecta in Figure~3 as measurements of the ambient wind rather than CME leg remnants.

\subsection{The Ejecta} 

Figure~4 presents an expanded view of the in situ measurements across the ejecta. The interval of the ejecta is identified by combining the enhancements in the alpha-to-proton density ratio, density and velocity. The density ratio (Figure~4b) from the shock to the leading portion of the ejecta is problematic and thus removed \citep[also see][]{romeo2023}. The value of the density ratio within the ejecta may not be accurate as some of the ion velocity distributions are outside the field of view or energy range of SPAN-I, but its enhancement can be used as an indicator of ejecta plasma. The stream with a decreasing velocity in the wake of the ejecta is a CME-induced ambient fast flow (Figure~4d). Another useful signature is the slightly depressed proton temperature in comparison with the expected one (Figure~4e). The leading edge of the ejecta is determined mainly from this signature. We obtain the expected temperature from a well-established relationship between the speed and temperature \citep[e.g.,][]{lopez1987, richardson1995} incorporating a distance gradient. The distance gradient ($r^{-1.1}$) is derived by attempting to match the observed temperature surrounding the ejecta. The ejecta interval is short, lasting only about 7 hr. The magnetic field does not show a large-scale rotation inside the ejecta (Figure~4h), with the first part almost radial (i.e., $\theta \sim 0^{\circ}$) and the middle part nearly perpendicular to the radial direction (i.e., $\theta \sim 90^{\circ}$). We see an indication of rotation only near the end of the ejecta. These likely indicate multiple structures within the ejecta. Indeed, there are multiple loops behind the shock in the imaging observations (Figure~1e). An intermittent BDE signature (Figure~4a) is observed only in the $\theta \sim 0^{\circ}$ portion. These characteristics are consistent with a crossing of the ejecta flank inferred from coronagraph images.

Note three unique, important features associated with the young ejecta compared to well-evolved ICMEs. First, the ambient wind is nearly sub-Alfv\'enic (i.e., $M_{\rm A}\sim1$) on the two sides of the event (Figure~4g), so the shock and ejecta were still in the coronal regime. It should not be called an ICME because it is not in interplanetary space yet. Second, the proton temperature inside the ejecta is significantly higher than that upstream of the shock. This indicates a hot solar source, and the ejecta is still at its early stage of expansion. This hot, pristine ejecta plasma agrees with the ``hot channel" in previous EUV observations \citep[e.g.,][]{zhang2012, cheng2013} and what the signature of charge state enhancement generally suggests \citep{lepri2004}. Third, the plasma $\beta$ is also high within the ejecta (higher than in the ambient medium), particularly in the trailing part of the ejecta ($\beta>1$). Note that the value of $\beta$ inside the ejecta should be considered as a lower limit because of the partial moments in the ion measurements. A large $\beta$ implies a highly non-force-free magnetic field structure. We evaluate the ratio of thermal pressure gradient to the Lorentz force, two major forces responsible for CME acceleration and expansion, as follows: ${\nabla p \over {\bf j}\times {\bf B}} \sim {p \over B^2/\mu} = {1\over 2}\beta$, where $\mu$ is the permeability constant. Here we have assumed that the length scales for the spatial variations of the magnetic field and thermal pressure are of the same order in magnitude. Therefore, a high $\beta$ also suggests that thermal pressure gradient is a crucial force in CME acceleration and expansion at the early stage. The early vast expansion often seen in large CMEs including the present one may have a significant contribution from this overpressure.      

\subsection{The Shock} 

Shown in Figure~5 is an expanded view of the in situ measurements across the coronal shock, which was observed at 17:27:19 UT on September 5 at a heliocentric distance of 15.06 solar radii. The shock arrival time is earlier than predicted by imaging observations. Note that downstream of the shock the peak of the QTN spectrum is contaminated by the associated electromagnetic emissions (Type II or III radio bursts). In order to minimize the uncertainty, we identify the frequency with the steepest slope right before the peak \citep[see][]{Meyer-Vernet2017, moncuquet2020}. This frequency is used as the local plasma frequency. The resulting density shows a peak right behind the shock (Figure~5a), which seems key in determining the shock parameters (see below). The velocity downstream of the shock is significantly non-radial (in particular the large $v_T$), so the shock motion at PSP is non-radial and the shock normal may have a substantial T component. 

We determine the shock parameters using a least-squares fit to the Rankine-Hugoniot conservation conditions \citep{vinas1986}. A reasonable fit can be obtained only when the density peak after the shock is chosen to represent the downstream density\footnote{If the time period corresponding to the density plateau behind the peak is chosen to represent the interval of the downstream asymptotic state, the downstream conditions are poorly fitted and some of the constants (e.g., the field component along the shock normal) are not conserved across the shock. In this case the fit yields a quasi-parallel shock and a shock velocity of about 1040 km s$^{-1}$ along the normal direction. However, these values are not trusted because of the poor fit.}. The density peak leads to a density compression ratio (2.5) across the shock comparable to the magnetic field compression ratio (2.3), so the shock is likely a quasi-perpendicular one. The shock is indeed quasi-perpendicular with an angle of about $104^{\circ}\pm5^{\circ}$ between the shock normal and the upstream magnetic field. The shock normal is $[0.34, -0.84, 0.43]$ in RTN coordinates, which agrees with our expectation from the velocity measurements. The shock is not particularly strong with an Alfv\'en Mach number of only about $2.1\pm0.1$. The shock velocity along the normal direction ($840\pm50$ km s$^{-1}$) is much smaller than inferred from imaging observations ($1300\pm 100$ km s$^{-1}$). 

We suggest that this is not the same shock as in the images, given the inconsistencies mentioned above. There could be another shock in the in situ data, as implied by the continuous increase of the speed toward the ejecta (Figure~5b), the peak speed in the sheath comparable to the shock velocity from imaging observations, and the two-step profiles of the velocity, temperature and field strength in the sheath. A careful examination of the data indeed indicates a discontinuity around 17:34 UT in the plasma and field parameters. The second shock is likely to have decayed: if we take the shock velocity of 1300 km s$^{-1}$ from imaging observations, the velocity of the shock relative to the medium upstream of it would be only about 300 km s$^{-1}$, which is significantly smaller than its upstream Alfv\'en speed (about 900 km s$^{-1}$). The two-shock scenario is consistent with more than one eruptions. The two shocks are expected to merge soon when the second one emerges from the sheath. SolO observed a shock passage at 10:01 UT on September 6 at a distance of about 0.7 au, which is likely the merged shock. Readers are directed to \citet{trotta2023} for a comparison of the shock properties at PSP and SolO. 

\subsection{The Reforming HCS}

To help interpret the measurements, we reconstruct the global coronal magnetic fields with a potential field source surface (PFSS) model \citep[e.g.,][]{altschuler1969, schatten1969, wang1992, badman2020}. Standard synoptic magnetograms from SDO/HMI \citep{hoeksema2014} are used as boundary conditions for the model, and the height of the source surface is set to 2.2 solar radii (the height is chosen based on the comparison between the modeled open field regions and the observed coronal holes in EUV images). The connectivity between PSP and the source surface is established with a Parker spiral field, whose curvature is determined by the solar wind radial velocity measured at PSP. The modeling results for Carrington rotations (CRs) 2261 - 2263 are displayed in Figure~6 as synoptic maps. A first impression about the results is a long-duration change in the morphology of the HCS. During CR 2262, the photospheric fields corresponding to the source active region (observed around August 23) are updated to a mature state (observed around September 19), and the HCS becomes more warped than before. The configuration does not recover to its previous state during CR 2263, so the change seems long-duration. This situation is similar to the case of \citet{liu2009}. Clearly, PSP would not cross any HCS during encounter 13, if it were configured as in CR 2261. However, we do see two polarity changes in the magnetic field from PSP measurements, which is consistent with the HCS configuration in CR 2262.

The first polarity change occurs inside the ejecta (Figure~4f), which is rather gradual with $B_R$ fluctuating around zero for a relatively sustained time. This gradual change instead of a definite HCS crossing is likely a consequence of the complex CME that was altering the field topology. The second polarity change is a clear HCS crossing (Figure~3f). This HCS crossing coincides with the prediction of PFSS modeling for CR 2262, but note that the HCS may slightly evolve between the source surface and the PSP distance (about 15 solar radii). Also shown in Figure~6 (CR 2262) is the magnetic connectivity for the sub-Alfv\'enic intervals around the HCS crossing. The sub-Alfv\'enic interval before the HCS crossing is connected to small areas of open magnetic fields (or small coronal holes) from the source active region. The origin from the active region may explain the higher solar wind density in the interval (Figure~3c). The sub-Alfv\'enic interval after the HCS crossing is connected to the boundary of an equatorial coronal hole, and the solar wind density is much lower in this case (Figure~3c). Both cases are consistent with their nature as LMBLs, where the rapidly diverging open magnetic fields result in low solar wind velocities \citep{liu2023}. The extremely low velocity (as low as 100 km s$^{-1}$) in the sub-Alfv\'enic interval after the HCS crossing (Figure~3d) is the primary contributor to the near crossing of the sonic critical point. Therefore, an LMBL may also provide favorable conditions for crossings of the sonic critical point as mentioned earlier. 

The HCS crossing is shown in Figure~7, and its geometry is illustrated in Figure~8. The field rotation angle is about 169$^{\circ}$ across the HCS. Reconnection signatures are observed inside the HCS, including the decreased field strength, increased radial velocity, and enhanced proton and electron temperatures (Figure~7). The enhancement in the radial velocity implies that the reconnection site (or X-line) is located sunward of PSP (Figure~8). The solar wind density (Figure~7g) is not enhanced inside the HCS as in other reconnection exhaust events \citep[e.g.,][]{gosling2005, chen2021, lavraud2021, phan2022}; perhaps the HCS is still in the process of reformation. We see abrupt changes in the radial component of the field near the edges (Figure~7c), which indicate kinks in the reconnected field lines \citep{petscheck1964}. The changes in $B_R$ and $v_R$ are anti-correlated upon entry and correlated upon exit of the HCS (Figure~7c and 7e), which is consistent with reconnection and suggests that the kinks are propagating away from the recognition site at the local Alfv\'en speed \citep{gosling2005}.  

We set up the current sheet coordinate system using a minimum variance analysis of the normalized magnetic field inside the HCS \citep{sonnerup1967}. The normal of the HCS (\textbf{Z}) should be along the minimum variance direction; the maximum variance would occur along the anti-parallel fields (\textbf{X}) since the field changes its sign across the HCS; the intermediate variance direction (\textbf{Y}) is identified as the X-line direction because of the nonuniform distribution of the guide field. The analysis yields $\textbf{X} = [0.996, -0.088, -0.034]$, $\textbf{Y} = [0.087, 0.996, -0.026]$, and $\textbf{Z} = [0.036, 0.022, 0.999]$, which are very close to the RTN coordinate system (Figure~8). The angular uncertainties of these vectors are estimated to be several degrees. The reconnection exhaust points along \textbf{X} that has non-negligible negative T and N components. Indeed, we observe enhanced negative $v_T$ and $v_N$ inside the HCS (Figure~7f). The HCS lies almost along the RT plane (Figure~8), but our PFSS modeling predicts a largely vertical orientation (Figure~6, CR 2262). Again, the HCS may still be in the process of reformation, or the discrepancy is due to uncertainties in the PFSS modeling. The HCS width can be estimated using the measurements, i.e., $d=({\bf v_{\rm sc}} - {\bf v_{\rm sw}}) t_{\rm d} \cdot {\bf Z}$, where ${\bf v_{\rm sc}}$ and ${\bf v_{\rm sw}}$ are the average velocities of the spacecraft ($[47, 146, 3]$ km s$^{-1}$) and solar wind ($[583, -133, -48]$ km s$^{-1}$) inside the HCS, and $t_{\rm d}$ is the time duration of the HCS (about 740 s). The resulting HCS width is about $(2.8\pm0.5)\times 10^4$ km, which is similar to other HCS widths in the near-Sun solar wind \citep{phan2022}.

Note that the HCS crossing and associated reconnection occur in the corona (i.e., the sub-Alfv\'enic regime) as deep as the sonic critical point (Figure~7i). If we draw an analogy between the HCS and a post-CME current sheet that is often discussed in the literature \citep[e.g.,][]{ko2003, lin2007}, we may expect a similar width for a post-CME current sheet in coronal conditions. The present HCS thickness corresponds to the lower end of the widths of post-CME current sheets estimated from remote-sensing observations \citep{lin2007}. We may also anticipate similar reconnection signatures in a post-CME current sheet, if the coronal conditions are not dramatically different. Another important implication is the consequence of coronal reconnections on the long-term balance of magnetic flux in the heliosphere \citep[e.g.,][]{gosling1975, mccomas1995, crooker2002}. On the opposite side of the X-line the reconnection outflow and reconnected field lines propagate toward the Sun (Figure~8), since the reconnection exhaust is in the sub-Alfv\'enic wind. If the solar wind were super-Alfv\'enic, they would be carried away from the Sun by the solar wind. Those reconnected field lines are able to return to the Sun and close down, thus reducing the amount of magnetic flux in the heliosphere. The returned plasma and closed field lines also seem necessary for the reformation of the streamer underneath the HCS (Figure~8).   

\section{Conclusions}

In this study we have examined the first direct in situ measurements of a fast CME and shock in the corona (i.e., the sub-Alfv\'enic regime), which occurred on 2022 September 5 when PSP was near the perihelion of encounter 13. Key findings are revealed concerning the structures of the large CME and shock at the early stage, the consequences of the CME to the corona and heliosphere, and coronal conditions at the sonic critical point. We summarize the results as follows.  

(1) Coronagraph imaging observations suggest that the CME may be composed of more than one eruptions. The complex CME shows a vast expansion at the early stage, producing a fast forward shock. The shock can be modeled well by a simple spherical structure. The modeling gives a peak speed of about 2900 km s$^{-1}$ and a propagation direction of about 170$^{\circ}$ west of the Earth and about 50$^{\circ}$ south of the ecliptic plane. PSP would encounter the flank of the ejecta and shock given the propagation direction with respect to the spacecraft. The model also predicts shock arrival at PSP around 17:39 UT on September 5 with a shock velocity of about 1300 km s$^{-1}$ along the normal direction.    

(2) The in situ characteristics of the ejecta are consistent with a flank crossing inferred from coronagraph images. The interval of the ejecta is identified from the enhancements in the alpha-to-proton density ratio, density and velocity, and the slightly depressed proton temperature in comparison with the expected one. The interval lasts only about 7 hr, and does not show a large-scale rotation in the magnetic field. The in situ measurements indicate three unique features associated with the young ejecta compared to well-evolved ICMEs: a nearly sub-Alfv\'enic environment; a proton temperature much higher than that of the ambient medium, suggestive of a hot solar source; a large plasma $\beta$, implying a highly non-force-free structure and the importance of thermal pressure gradient for CME acceleration and expansion.

(3) The shock from in situ measurements around 17:27 UT on September 5 should not be the same shock as in the images. The shock velocity along the normal direction (about 840 km s$^{-1}$) is much smaller than inferred from coronagraph images (about 1300 km s$^{-1}$), and the shock arrival time is earlier than predicted. The shock is not particularly strong with an Alfv\'en Mach number of only about 2.1. It is a quasi-perpendicular shock with an angle of about 104$^{\circ}$ between its normal and the upstream magnetic field. We identify a discontinuity around 17:34 UT on September 5 in the plasma and field parameters, which is likely a decayed shock. The second shock has decayed, because it was propagating in a fast flow with a large Alfv\'en speed (i.e., the sheath of the first shock). The two-shock scenario is consistent with more than one eruptions. We suggest that the second shock corresponds to the stronger eruption (if there are indeed two successive eruptions). The two shocks are expected to merge soon when the second one emerges from the sheath. 

(4) The HCS becomes more warped than before. This seems a long-duration change, but the HCS in PSP in situ measurements may not reach the mature state yet. In situ measurements of the magnetic field indicate two polarity reversals, which agrees with the more warped HCS configuration. The first polarity reversal occurs inside the ejecta and is rather gradual. The second one is a clear HCS crossing, with reconnection signatures as deep as the sonic critical point. The HCS crossing has a width of about $2.8\times 10^4$ km in the corona. These results have important implications for post-CME current sheets and the long-term balance of magnetic flux in the heliosphere. We may expect a similar width and similar reconnection signatures for a post-CME current sheet, if the coronal conditions are not dramatically different. Since the reconnection occurs in the sub-Alfv\'enic wind, the reconnected field lines sunward of the reconnection site are able to return to the Sun and close down, which helps balance magnetic flux in the heliosphere. 

(5) We also obtain results important for understanding coronal conditions at the sonic critical point and the origin and evolution of switchbacks. The way that ADs (our term for switchbacks to include small deflections) vary in the measurements outside the transient plasma corresponds well with the Alfv\'en Mach number, which is consistent with the theory of \citet{liu2023}. Our magnetic mapping indicates that the sub-Alfv\'enic intervals around the HCS crossing are LMBLs defined by \citet{liu2023}. The extremely low velocity (as low as 100 km s$^{-1}$) in the second sub-Alfv\'enic interval leads to in situ measurements of the coronal plasma deep to the sonic critical point. It implies a surprisingly large sonic critical radius of about 15 solar radii. An LMBL may provide favorable conditions for the crossings of the sonic critical point in addition to the Alfv\'en surface. Based on the measurements near the sonic critical point, we suggest that ADs may start from the sonic critical point, as below it the magnetic field is too strong to be deflected.

\acknowledgments The research was supported by the Strategic Priority Research Program of the Chinese Academy of Sciences (No. XDB0560000), NSFC under grants 42274201 and 42004145, the National Key R\&D Program of China (No. 2021YFA0718600 and No. 2022YFF0503800), and the Specialized Research Fund for State Key Laboratories of China. We acknowledge the NASA Parker Solar Probe mission and the SWEAP and FIELDS teams for use of data. The PFSS extrapolation is performed using the \emph{pfsspy} Python package \citep{stansby2020}. The data used for PFSS modeling are courtesy of SDO.

\clearpage

\begin{figure}
\epsscale{1.0} \plotone{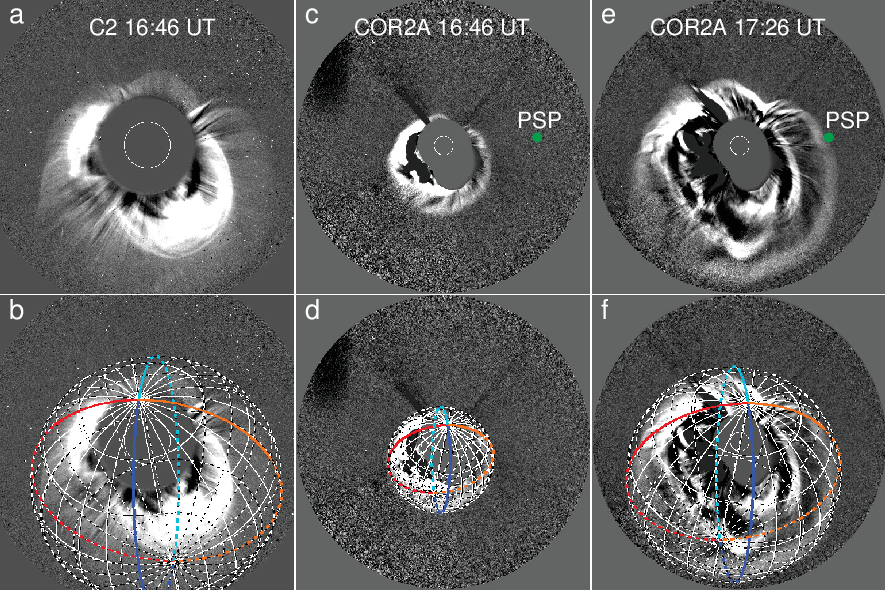} 
\caption{Running-difference coronagraph images of the CME and shock on September 5. (a-b) Image from LASCO C2 of SOHO and corresponding shock modeling at 16:46 UT. (c-d) Image from COR2 of STEREO A and corresponding shock modeling at 16:46 UT. (e-f) Image from COR2 of STEREO A and corresponding shock modeling at 17:26 UT. The projected location of PSP is indicated on the COR2 images.} 
\end{figure}

\clearpage

\begin{figure}
\epsscale{0.9} \plotone{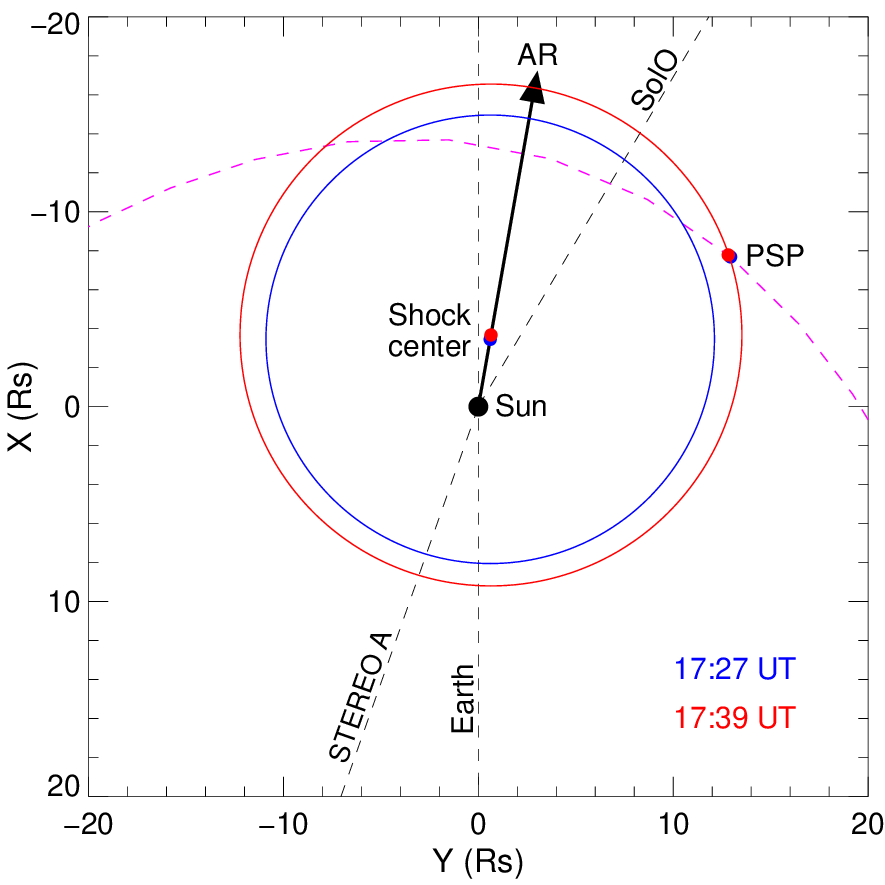} 
\caption{Cross sections of the modeled shock in the ecliptic plane at two different times on September 5. The directions of the Earth, STEREO A (19.4$^{\circ}$ east), and SolO (149.4$^{\circ}$ west) are indicated by the dashed lines. The pink curve is the trajectory of PSP. The black arrow marks the longitude of the active region (AR), which is the propagation direction of the shock in the ecliptic plane. Also shown are the location of PSP and the center of the spherical shock. PSP was about 121$^{\circ}$ west of the Earth and about 15 solar radii from the center of the Sun at the two times, although it slightly moved.}
\end{figure}

\clearpage

\begin{figure}
\epsscale{0.85} \plotone{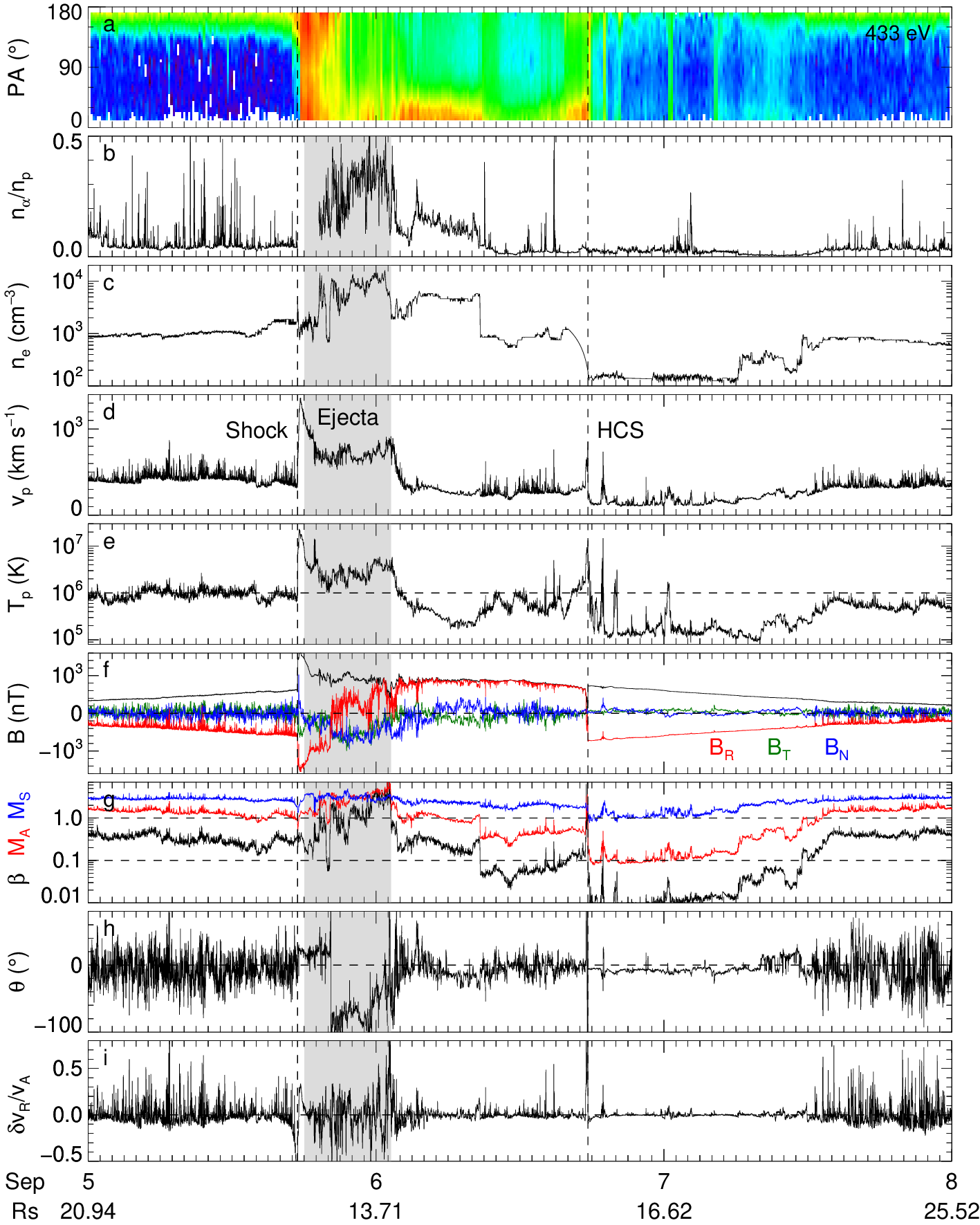} 
\caption{Overview of PSP measurements at encounter 13. (a) Pitch angle (PA) distribution of 433 eV electrons. (b) Alpha-to-proton density ratio. (c) Electron density from QTN. (d) Bulk speed. (e) Proton temperature. (f) Magnetic field strength and components. (g) Plasma $\beta$, radial Alfv\'en Mach number, and radial sonic Mach number. (h) Magnetic field deflection angle. (i) Radial velocity variation in units of local Alfv\'en speed. The shaded region indicates the ejecta interval. The vertical dashed lines mark the preceding shock and the HCS crossing, respectively.}
\end{figure}

\clearpage

\begin{figure}
\epsscale{0.9} \plotone{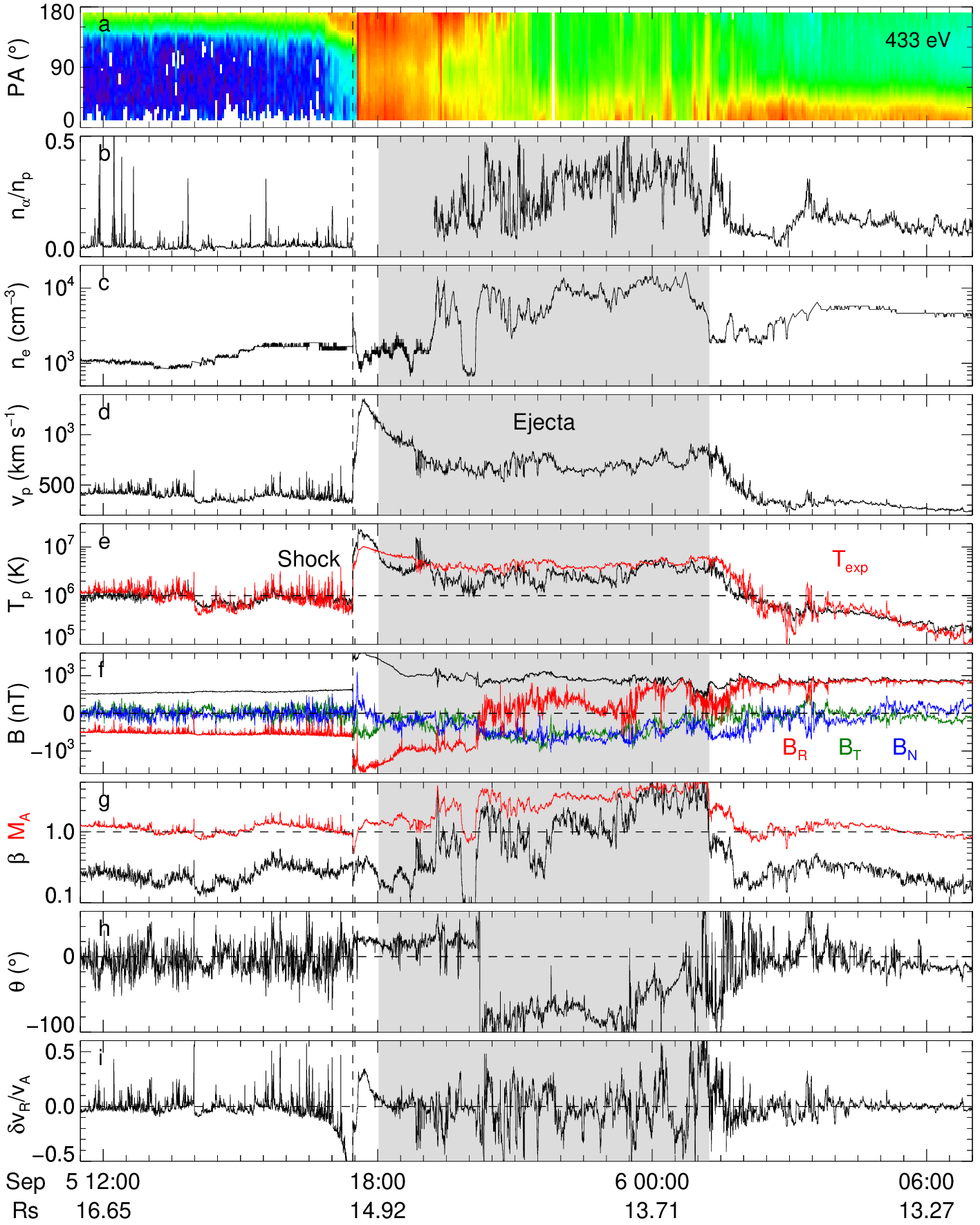} 
\caption{Expanded view of the measurements across the ejecta. Similar to Figure~3. The red curve in panel (e) denotes the expected proton temperature calculated from the observed speed.}
\end{figure}

\clearpage

\begin{figure}
\epsscale{0.8} \plotone{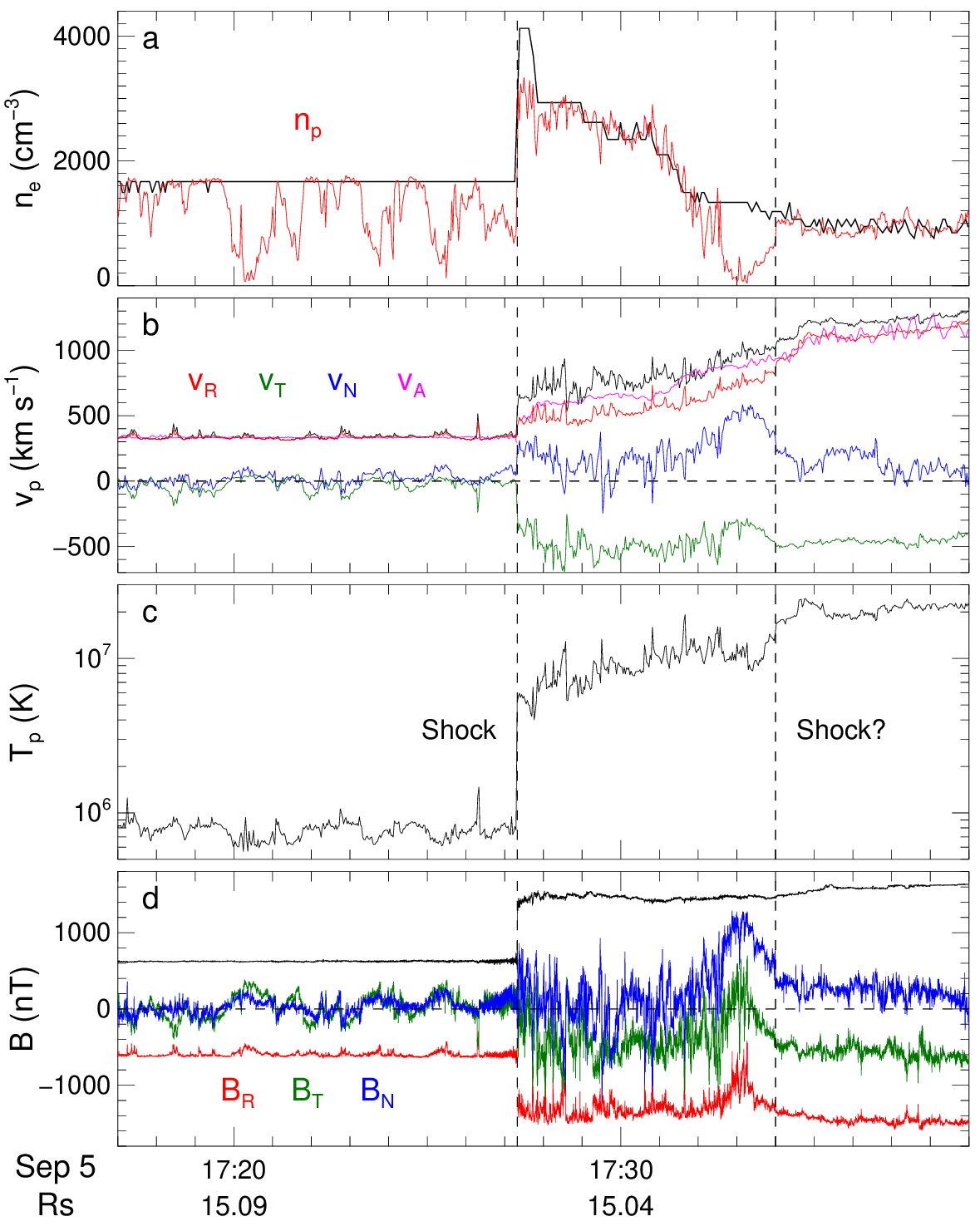} 
\caption{Expanded view of the measurements across the shock. Similar to Figure~3. Here we add the proton density from SPAN-I, velocity components, and the Alfv\'en speed. The second vertical dashed line marks a discontinuity in the plasma and field parameters, which is likely a decayed shock.}
\end{figure}

\clearpage

\begin{figure}
\epsscale{0.7} \plotone{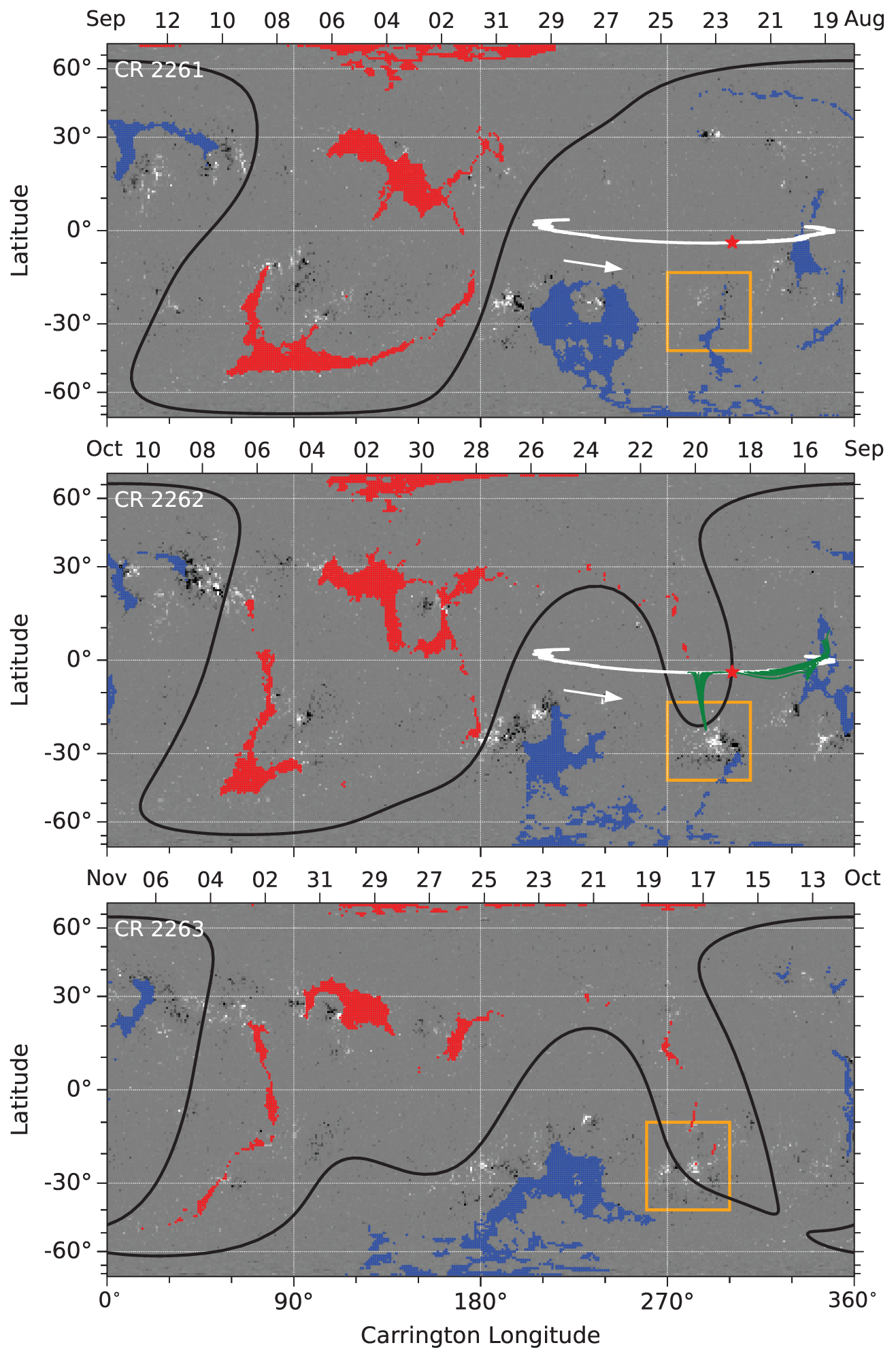} 
\caption{PFSS model results projected onto the synoptic maps of the photospheric field for CRs 2261 - 2263. The black curve is the source surface neutral line (the coronal base of the HCS). The red and blue areas indicate the footpoints of positive and negative open fields, respectively. The white curve is the trajectory of PSP (September 1 - 10) projected onto the source surface, and the white arrow shows the direction of motion of the spacecraft. The red star marks the location of PSP at the time of the in situ HCS crossing. The green lines in the middle panel represent the magnetic connectivity of the sub-Alfv\'enic intervals from the PSP trajectory to the photospheric sources. The orange square indicates the position of NOAA AR 13088. The dates corresponding to each CR are given at the top.}
\end{figure}

\clearpage

\begin{figure}
\epsscale{0.9} \plotone{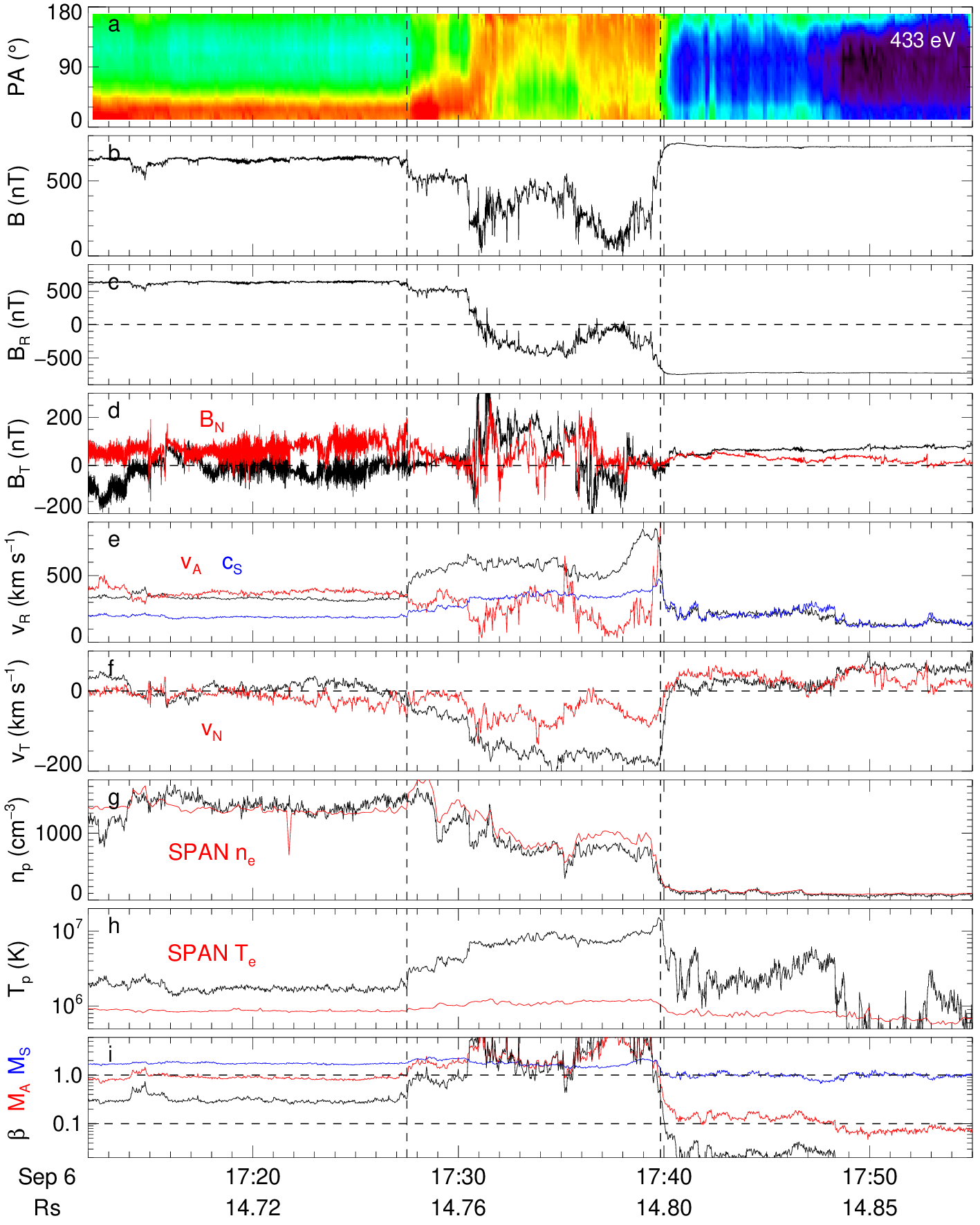} 
\caption{Expanded view of the measurements across the HCS. Similar to Figures~3 and 5. The vertical dashed lines mark the edges of the reconnection exhaust. Here we add the sound speed ($c_{\rm s}$), and the electron density and temperature from SPAN-E. The electron density from QTN is not shown here because of a data gap around the HCS.}
\end{figure}

\clearpage

\begin{figure}
\epsscale{0.9} \plotone{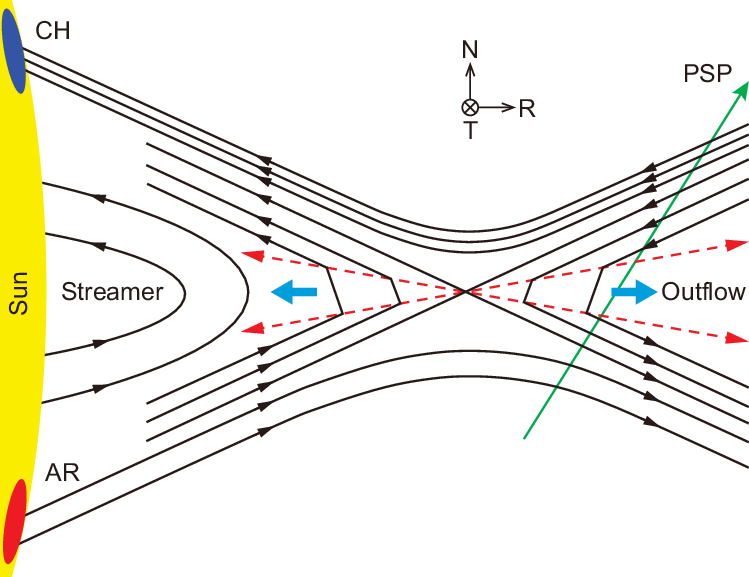} 
\caption{Schematic illustration of the reforming streamer, HCS and associated reconnection exhaust below the Alfv\'en critical point. The open field lines of opposite polarities are from the coronal hole (CH) and the active region (AR) according to the magnetic mapping. The reconnection is asymmetric with stronger fields from the CH. The HCS lies almost along the RT plane. PSP passes through the exhaust region anti-sunward of the reconnection site, with an average velocity of [47, 146, 3] km s$^{-1}$ in RTN coordinates. The field line kinks associated with newly merged field lines propagate sunward and anti-sunward from the reconnection site at the respective Alfv\'en speeds there. The dashed red arrows, which pass through the kinks, define the boundaries of the exhaust regions. The reconnected field lines sunward of the reconnection site are connected to the Sun (not drawn here).}
\end{figure}


\begin{thebibliography}{}

\bibitem[Altschuler \& Newkirk(1969)]{altschuler1969}
Altschuler, M. D., \& Newkirk, G. 1969, SoPh, 9, 131

\bibitem[Badman et al.(2020)]{badman2020}
Badman, S. T., Bale, S. D., Mart\'{\i}nez Oliveros, J. C., et al. 2020, ApJS, 246, 23

\bibitem[Bale et al.(2016)]{bale2016}
Bale, S. D., Goetz, K., Harvey, P. R., et al. 2016, SSRv, 204, 49 

\bibitem[Bale et al.(2019)]{bale2019}
Bale, S. D., Badman, S. T., Bonnell, J. W., et al. 2019, Natur, 576, 237

\bibitem[Burlaga et al.(1981)]{burlaga1981}
Burlaga, L. F., Sittler, E., Mariani, F., \& Schwenn, R. 1981, JGR, 86, 6673

\bibitem[Burlaga(1988)]{burlaga1988}
Burlaga, L. F. 1988, JGR, 93, 7217

\bibitem[Chen et al.(2021)]{chen2021}
Chen, C., Liu, Y. D., \& Hu, H. 2021, ApJ, 921, 15

\bibitem[Cheng et al.(2013)]{cheng2013}
Cheng, X., Zhang, J., Ding, M. D., Liu, Y., \& Poomvises, W. 2013, ApJ, 763, 43 

\bibitem[Crooker et al.(2002)]{crooker2002}
Crooker, N. U., Gosling, J. T., \& Kahler, S. W. 2002, JGR, 107, 1028

\bibitem[DeForest et al.(2014)]{deforest2014}
DeForest, C. E., Howard, T. A., \& McComas, D. J. 2014, ApJ, 787, 124

\bibitem[Forbes(2000)]{forbes2000}
Forbes, T. G. 2000, JGR, 105, 23153

\bibitem[Fox et al.(2016)]{fox2016}
Fox, N. J., Velli, M. C., Bale, S. D., et al. 2016, SSRv, 204, 7  

\bibitem[Goldstein(1983)]{goldstein1983}
Goldstein, H. 1983, in Solar Wind Five, ed. M. Neugebauer (Washington, DC: NASA), 731

\bibitem[Gosling(1975)]{gosling1975}
Gosling, J. T. 1975, RvGSP, 13, 1053

\bibitem[Gosling et al.(2005)]{gosling2005}
Gosling, J. T., Skoug, R. M., McComas, D. J., \& Smith, C. W. 2005, JGR, 110, A01107

\bibitem[Halekas et al.(2020)]{halekas2020}
Halekas, J. S., Whittlesey, P., Larson, D. E., et al. 2020, ApJS, 246, 22

\bibitem[Hess \& Zhang(2014)]{hess2014}
Hess, P., \& Zhang, J. 2014, ApJ, 792, 49

\bibitem[Hoeksema et al.(2014)]{hoeksema2014}
Hoeksema, J. T., Liu, Y., Hayashi, K., et al. 2014, SoPh, 289, 3483

\bibitem[Hu et al.(2019)]{hu2019}
Hu, H., Liu, Y. D., Zhu, B., et al. 2019, ApJ, 878, 106

\bibitem[Jiao et al.(2024)]{jiao2024}
Jiao, Y., Liu, Y. D., Ran, H., \& Cheng, W. 2024, ApJ, 960, 42

\bibitem[Kasper et al.(2016)]{kasper2016}
Kasper, J. C., Abiad, R., Austin, G., et al. 2016, SSRv, 204, 131   

\bibitem[Kasper et al.(2019)]{kasper2019}
Kasper, J. C., Bale, S. D., Belcher, J. W., et al. 2019, Natur, 576, 228

\bibitem[Kasper et al.(2021)]{kasper2021}
Kasper, J. C., Klein, K. G., Lichko, E., et al. 2021, PhRvL, 127, 255101

\bibitem[Ko et al.(2003)]{ko2003}
Ko, Y.-K., Raymond, J. C., Lin, J., et al. 2003, ApJ, 594, 1068

\bibitem[Kwon et al.(2014)]{kwon2014}
Kwon, R., Zhang, J., \& Olmedo, O. 2014, ApJ, 794, 148

\bibitem[Kwon et al.(2015)]{kwon2015}
Kwon, R., Zhang, J., \& Vourlidas, A. 2015, ApJL, 799, L29

\bibitem[Lavraud et al.(2021)]{lavraud2021}
Lavraud, B., Kieokaew, R., Fargette, N., et al. 2021, A\&A, 656, A37

\bibitem[Lepping et al.(1990)]{lepping1990}
Lepping, R. P., Jones, J. A., \& Burlaga, L. F. 1990, JGR, 95, 11957

\bibitem[Lepri \& Zurbuchen(2004)]{lepri2004}
Lepri, S. T., \& Zurbuchen, T. H. 2004, JGR, 109, A01112

\bibitem[Lin et al.(2007)]{lin2007}
Lin, J., Li, J., Forbes, T. G., Ko, Y.-K., Raymond, J. C., \& Vourlidas, A. 2007, ApJL, 658, L123

\bibitem[Liu et al.(2006)]{liu2006}
Liu, Y., Richardson, J. D., Belcher, J. W., Kasper, J. C., \& Elliott, H. A. 2006, JGR, 111, A01102

\bibitem[Liu et al.(2009)]{liu2009}
Liu, Y., Luhmann, J. G., Lin, R. P., et al. 2009, ApJL, 698, L51 

\bibitem[Liu et al.(2017)]{liu2017}
Liu, Y. D., Hu, H., Zhu, B., Luhmann, J. G., \& Vourlidas, A. 2017, ApJ, 834, 158

\bibitem[Liu et al.(2019a)]{liu2019a}
Liu, Y. D., Zhu, B., \& Zhao, X. 2019a, ApJ, 871, 8 

\bibitem[Liu et al.(2019b)]{liu2019b}
Liu, Y. D., Zhao, X., Hu, H., Vourlidas, A., \& Zhu, B. 2019b, ApJS, 241, 15

\bibitem[Liu et al.(2021)]{liu2021}
Liu, Y. D., Chen, C., Stevens, M. L., \& Liu, M. 2021, ApJL, 908, L41 

\bibitem[Liu et al.(2023)]{liu2023}
Liu, Y. D., Ran, H., Hu, H., \& Bale, S. D. 2023, ApJ, 944, 116 

\bibitem[Livi et al.(2022)]{livi2022}
Livi, R., Larson, D. E., Kasper, J. C., et al. 2022, ApJ, 938, 138

\bibitem[Long et al.(2023)]{long2023}
Long, D. M., Green, L. M., Pecora, F., et al. 2023, ApJ, in press   

\bibitem[Lopez(1987)]{lopez1987}
Lopez, R. E. 1987, JGR, 92, 11189

\bibitem[Low(1996)]{low1996}
Low, B. C. 1996, SoPh, 167, 217

\bibitem[Low(2001)]{low2001}
Low, B. C. 2001, JGR, 106, 25141

\bibitem[Matthaeus(2021)]{matthaeus2021}
Matthaeus, W. H. 2021, eprint arXiv: 2106.08450, doi: 10.48550/arXiv.2106.08450 

\bibitem[McComas(1995)]{mccomas1995}
McComas, D. J. 1995, RvGSP, 33, 603

\bibitem[Meyer-Vernet et al.(2017)]{Meyer-Vernet2017}
Meyer-Vernet, N., Issautier, K., \& Moncuquet, M. 2017, JGR, 122, 7925

\bibitem[Moncuquet et al.(2020)]{moncuquet2020}
Moncuquet, M., Meyer-Vernet, N., Issautier, K., et al. 2020, ApJS, 246, 44 

\bibitem[Neugebauer \& Goldstein(1997)]{neugebauer1997}
Neugebauer, M., \& Goldstein, R. 1997, in Geophys. Monogr. Ser. 99, Coronal Mass Ejections, ed. N. Crooker, J. A. Joselyn, \& J. Feynman (AGU, Washington, D. C.), 245

\bibitem[Paouris et al.(2023)]{paouris2023}
Paouris, E., Vourlidas, A., Kouloumvakos, A., et al. 2023, ApJ, 956, 58

\bibitem[Patel et al.(2023)]{patel2023}
Patel, R., West, M. J., Seaton, D. B., et al. 2023, ApJL, 955, L1

\bibitem[Petscheck(1964)]{petscheck1964}
Petscheck, H. E. 1964, NASSP, 50, 425

\bibitem[Phan et al.(2022)]{phan2022}
Phan, T. D., Verniero, J. L., Larson, D., et al. 2022, GeoRL, 49, e96986 

\bibitem[Raouafi et al.(2023)]{raouafi2023}
Raouafi, N. E., Matteini, L., Squire, J., et al. 2023, SSRv, 219, 8

\bibitem[Richardson \& Cane(1995)]{richardson1995}
Richardson, I. G., \& Cane, H. V. 1995, JGR, 100, 23397

\bibitem[Romeo et al.(2023)]{romeo2023}
Romeo, O. M., Braga, C. R., Badman, S. T., et al. 2023, ApJ, 954, 168

\bibitem[Schatten et al.(1969)]{schatten1969}
Schatten, K. H., Wilcox, J. M., \& Ness, N. F. 1969, SoPh, 6, 442

\bibitem[Sonnerup \& Cahill(1967)]{sonnerup1967}
Sonnerup, B. U. \"O, \& Cahill, L. J., Jr. 1967, JGR, 72, 171

\bibitem[Stansby et al.(2020)]{stansby2020}
Stansby, D., Yeates, A., \& Badman, S. 2020, JOSS, 5, 2732

\bibitem[Trotta et al.(2023)]{trotta2023}
Trotta, D., Larosa, A., Nicolaou, G., et al. 2023, ApJ, submitted 

\bibitem[Vi\~{n}as \& Scudder(1986)]{vinas1986}
Vi\~{n}as, A. F., \& Scudder, J. D. 1986, JGR, 91, 39

\bibitem[Wang \& Sheeley(1992)]{wang1992}
Wang, Y.-M., \& Sheeley, N. R., Jr. 1992, ApJ, 392, 310

\bibitem[Weber \& Davis(1967)]{weber1967}
Weber, E. J., \& Davis, L. Jr. 1967, ApJ, 148, 217

\bibitem[Whittlesey et al.(2020)]{whittlesey2020}
Whittlesey, P. L., Larson, D. E., Kasper, J. C., et al. 2020, ApJS, 246, 74

\bibitem[Zhang et al.(2012)]{zhang2012}
Zhang, J., Cheng, X., \& Ding, M. D. 2012, NatCo, 3, 747

\bibitem[Zhu et al.(2018)]{zhu2018}
Zhu, B., Liu, Y. D., Kwon, R.-Y., \& Wang, R. 2018, ApJ, 865, 138

\bibitem[Zurbuchen \& Richardson(2006)]{zurbuchen2006}
Zurbuchen, T. H., \& Richardson, I. G., 2006, SSRv, 123, 31

\end{thebibliography}
\end{document}